\documentclass[twocolumn,twoside,slac]{revtex4}
\usepackage{graphicx}
\usepackage{fancyhdr}
\begin{document}

\title{Grid-based access control for Unix environments, Filesystems and Web
 Sites}

%

\author{A. McNab}
\affiliation{Schuster Laboratory, 
University of Manchester, Manchester, M13 9PL, United Kingdom}

\begin{abstract}
The EU DataGrid has deployed a grid testbed at approximately 20 sites across
Europe, with several hundred registered users. This paper describes
authorisation systems produced by GridPP and currently used on the EU
DataGrid Testbed,
including local Unix pool accounts and fine-grained access
control with Access Control Lists and Grid-aware filesystems, fileservers
and web developement environments.  

\end{abstract}

\maketitle

\thispagestyle{fancy}


\section{Introduction}

The overall security infrastructure of the EU DataGrid (EDG)\cite{edg}
project has been
described elsewhere\cite{SCGCHEP}. In brief, it consists of Certificate
Authorities (CA) granting X.509 cryptographic certificates to users and hosts,
and providing authentication; and Virtual Organisations (VO) which authorize
users to use resources allocated to the VO. Authorization information is
currently published as VO membership lists using LDAP, but VOMS, a system
based on attribute certificates has been developed and
described\cite{SCGCHEP}.

This paper describes ways in which this authorization information is used to
control access to local resources, such as Unix accounts, disk filesystems and
the virtual filesystems exported by file and web servers.

\section{Unix Account Management}

The remote job execution middleware used by EDG is based on the
Globus\cite{globus} gatekeeper, and this uses a static mapping from grid
identities, based on X.509 certificates, to local Unix user accounts.
A text file lists grid identities and corresponding local accounts and when
a job is received for execution, it is forked as a process owned by the
local Unix user account. Whilst adaquate for small Virtual Organisations,
this procedure is insufficient for VOs with hundreds of users at tens of
sites. To address the maintainence of the list of acceptable grid
identities, EDG has developed software to allow VOs to publish membership
lists and for sites to construct the mapping text file\cite{SCGCHEP}.

However, this still leaves the creation and management of the local Unix
accounts themselves. To address this for EDG, we have developed
a system of dynamically allocated Pool Accounts\cite{gppauthz},
which are created by the site administrator and
then allocated to users as new job or file server requests are received. A
directory of lock files is maintained to retain a one-to-one mapping between
grid identities and allocated user accounts. This also ensures that if two
requests overlap, they are assigned to the same Unix user account, which
allows sharing of files between multipart jobs, and between jobs and
fileserver requests.

Since Unix operating systems implement filesystem permissions in terms of
Unix accounts, this also provides a rudimentary way of preventing jobs from
different grid users from interfering with each other.

Accounts may be returned to the pool of unused accounts once all jobs
running as that account have terminated and after leaving a grace period for
file retrieval. Since the account allocation book keeping is maintained by
lock files, this is straightforward to implement as a Unix shell script
which is run periodically and which can be tailored to individual site
requirements.

\section{Grid Access Control Lists}

To describe fine-grained access control of files and other file-like
resources, EDG has developed GACL\cite{gppauthz},
a format for access control lists,
written in XML and in terms of grid identities or Virtual Organisation
membership. 

Each GACL access control list is divided into one or more entries, each of
which has a set of permissions which are granted if that entry's
credential requirements are met. Permissions are to Read (to read files), 
List (to obtain directory listings), Write (to create or write to files, to
create directories, or to delete files or directories) and
Admin (to modify access control lists.)

An entry may have one or more credentials which must be present, including
X.509 certificate identities, VO groups or VOMS attribute
certificates. Two generic credentials, Authuser (any user with a valid
certificate) and Anyuser (any user irrespective of credentials), allow access
to be granted to users with no affiliation to the site.

An API and library are provided for manipulating GACL lists, and this is the
foundation of the filesystem and fileserver access control described in the
remainder of this paper, and of the EU DataGrid Storage Element described
elsewhere\cite{se}.

\section{SlashGrid filesystems}

We have paid particular attention to applying the GACL access control to
standard local filesystem operations, using the SlashGrid\cite{gppauthz}
framework described here.

Most applications use a filesystem interface to access local files on
the same machine. This organises data into files, contained in a hierarchy
of folders or directories, each accessible by name. For interactive use, a
graphical file browser is commonly used, displaying files as icons which may
be opened and accessed using a mouse. File access within an applications
uses an analogous programming interface, which in most programming languages
is based on a set of functions to `open', `read', `write' etc.
 
The security associated with these operations is traditionally tied to
credentials which only have meaning on the machine (or in some cases the
computing site or cluster) in question. Typically, this takes the form of a
short username or group name, and a specific file may have one user who has
permission to write to that file. For EDG testbed sites, these are
dynamically allocated pool accounts, but there may be static accounts in other
parts of the system, such as user-interface hosts.

As we connect machines and sites together with Grid technology,
these local credentials become increasingly inappropriate for managing
authorisation to use resources, as they cannot readily be shared across the
Grid. For example, a user may have the username mcnab at one site, but
amcnab at another, and at a third site user mcnab may be a completely
different individual.

Although the pool accounts system described above automates the management
of local accounts, the
system cannot readily be used when creating long lived files, since the
username they are owned by is only temporarily associated with a specific
Grid identity.

Initially to resolve this shortcoming, we have produced a file system 
framework,
SlashGrid, which allows file and directory authorisation to depend on
long-lived Grid identities. 
SlashGrid creates a hierarchy of directories under /grid where an
application's username, whether static or temporary, is irrelevant to
whether it can create, read or modify files: what matters are the Grid
credentials the application currently holds on behalf of the user, wherever
they are on the Grid.

For interoperability with other products of the EU DataGrid and related
projects, SlashGrid uses the GACL library and access control lists stored in
per-directory or per-file control files.

SlashGrid has also been designed to be readily extensible, by the use of
third-party plugins to add additional filesystem types. In particular, we
have implemented an HTTP/HTTPS filesystem, in which the contents of remote
websites can be accessed by applications as if they were local files, and in
the case of HTTPS, may prove the user's identity to remote servers to obtain
access to restricted files.

This has the potential to allow existing applications to operate on
the Grid, indifferent to the true location of the files they manipulate,
with remote Grid file access provided as a service by the operating system
layer.

\section{GridSite web and file servers}

Web browsers represent the most common, familiar and most widely installed
application used to access remote resources on the current Internet.
However, most websites are built using HTTP technology, which can only
implement cumbersome authentication and authorisation mechanisms. Typically,
this involves the user choosing a short memorable password for each site to
which they need to identify themselves. Consequently, the user may find
themselves having to enter multiple usernames and passwords as they pass
between websites run by their employer, their bank, online merchants etc. As
well as the inconvenience involved, this is also vulnerable to ``brute
force'' attacks by third parties due to the short length of the passwords.

Since the mid-1990's, most web browsers have also supported  the HTTPS
protocol, which uses X.509 digital certificates and has been widely used to
provide authentication of websites to users. This allows a user to send
credit card details to a merchant's website, for instance, with some
confidence that the site is not being impersonated by a malicious third
party. 

Although the corresponding user authentication to websites has been
supported since the adoption of HTTPS, it has been far less used, due to the
administrative overhead and cost of verifying users' identity before giving
them a meaningful X.509 user certificate. 

However, with the large-scale deployment of X.509 certificates to members EDG
and other Grid projects across the world, this is changing, and it is
now practical to base a High Energy Physics collaboration's website on HTTPS
rather than HTTP
technology, without requiring users to install any special software.

The GridPP project, which represents the UK involvement in EDG,
has chosen to implement its collaboration website in this way,
and to produce a general website management tool, GridSite\cite{gppauthz},
which is
flexible enough for other projects to use for their own sites.

Since GridSite is able to uniquely and securely identify users by their X.509
certificate, they can be granted rights to edit and upload webpages, images
and binary files. This is enforced using the GACL access control lists
described above. Access control can be specified in terms of individuals or
Virtual Organisation groups, with membership managed by the group's
administrators through the same web interface.

This has allowed GridPP to devolve maintenance of the website down to the
level of those directly involved in each area of work. Since the
administration of group authorisation is also devolved, the administrative
overhead normally carried by the website manager is greatly reduced.

Since GridSite permits several users to maintain a set of documents, this
has also made collaboration between GridPP members at different institutions
considerably easier; and tools are provided to retain old versions and
record document histories to automate the book-keeping of who has changed a
document and at what date.

Initially, GridSite has been implemented as a self-contained executable run
from the web server to handle each HTTP or HTTPS request. It has now been
divided into standalone executables to handle interactive management of the
site and groups by administrators, and a loadable module which is
dynamically linked directly into the Apache\cite{apache} webserver used. By
incorporating GridSite and GACL technology directly in the webserver, all
technologies support by the webserver, including static file serving, and
dynamic content provided by CGI scipts, PHP, ASP or JSP server-parsed pages,
for example, can be subject to grid-based access control.

This flexibility
allows a GridSite server to simultaneously operate as an efficient file
server, as a web host with dynamic content and as a grid host with Grid
Services in Java and other languages operating in their favoured
environments.

\section{Future Developments}

Future developments will include porting of the SlashGrid filesystems
framework and the GridSite server management system to platforms
other than Linux. Support for additional authorization credentials, such as
Globus CAS, will be added to the GACL library, along with support for access
control languages recommended by the GGF Authorization Working Group.

\begin{acknowledgments}
This work has been
supported by the UK Particle Physics and Astronomy Research Council
and the European Union.
\end{acknowledgments}


\begin{thebibliography}{9}   

\bibitem {edg}
The EU DataGrid Project, \\ http://www.eu-datagrid.org/

\bibitem {SCGCHEP}
R. Alfieri et al. (EDG Security Co-ordination Group),
``Managing Dynamic User Communities in a Grid of Autonomous Resources'',
Proceedings of Computing in High Energy and Nuclear Physics (2003)

\bibitem {globus}
The Globus Project, \\ http://www.globus.org/

\bibitem {gppauthz}
GridPP access control tools: Pool Accounts, GACL, GridSite and SlashGrid,\\
http://www.gridpp.ac.uk/authz/

\bibitem{se}
The EU DataGrid Storage Element, \\
http://web01.esc.rl.ac.uk/projects/DataGrid/wp5/

\bibitem{apache}
The Apache Software Foundation, \\
http://www.apache.org/


\end{thebibliography}

\end{document}